\documentstyle[psfig,aps,twocolumn,graphicx,amsfonts,enumerate,epsfig]{revtex}  
\tighten
\begin{document}

\draft

\title{\bf Symmetries and Triplet Dispersion in a Modified Shastry-Sutherland Model
  for SrCu$_2$(BO$_3$)$_2$} 

\author{Christian Knetter\thanks{
e-mail: ck@thp.uni-koeln.de
\hspace*{\fill} {\protect\linebreak}
Internet: www.thp.uni-koeln.de/\~{}ck/}, Erwin M\"uller-Hartmann and G\"otz S. 
Uhrig\thanks{e-mail: gu@thp.uni-koeln.de
\hspace*{\fill} {\protect\linebreak}
Internet: www.thp.uni-koeln.de/\~{}gu/}}

\address{Institut f\"ur Theoretische Physik, Universit\"at zu
  K\"oln, Z\"ulpicher Str. 77, D-50937 K\"oln, Germany\\[1mm]
  {\rm(\today)} }

\maketitle

\begin{abstract}
We investigate the one-triplet dispersion in a modified
Shastry-Sutherland Model for SrCu$_2$(BO$_3$)$_2$ by means of a series
expansion about the limit of strong dimerization. Our perturbative
method is based on a continuous unitary transformation that maps the
original Hamiltonian to an effective, energy quanta conserving block
diagonal Hamiltonian ${\cal H}_{\rm eff}$. The dispersion splits into
two branches which are 
nearly degenerated. We analyse the symmetries of the model and
show that space group operations are necessary to explain the
degeneracy of the dispersion at ${\mathbf k}={\mathbf 0}$ and at the
border of the magnetic Brillouin zone. Moreover, we investigate the
behaviour of the dispersion for small $|{\mathbf k}|$ and compare our
results to INS data.
\end{abstract}


\narrowtext
\section{Introduction}
\begin{figure}[ht]
\begin{center}
{\psfig{figure=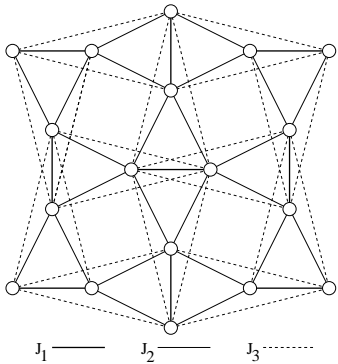 ,height=48mm}}
\end{center}
\caption{\footnotesize{An extract of the lattice we suggest for
    SrCu$_2$(BO$_3$)$_2$. The coupling $J_1$ is assumed to be
    antiferromagnetic. The starting point of our analysis is the limit
    of strong dimerization ($J_2,J_3 \rightarrow 0$).} \label{model}}
\end{figure}

The study of quantum spin systems exhibiting a finite spin gap has
advanced significantly through the recent synthesis of novel magnetic
materials. While  
quasi-one dimensional systems have been studied for a long time now,
new interest arose from the discovery of new quasi-two dimensional
materials like CaV$_4$O$_9$ \cite{tanig95} and (VO$_2$)P$_2$O$_7$ 
\cite{garre97a}. Particularly
interesting is SrCu$_2$(BO$_3$)$_2$ \cite{kagey99} since it is an
experimental realization of the Shastry-Sutherland model
\cite{miyah98,shast81b}. In Ref. \cite{mulle00} we presented an extended
version of that model (Fig.~\ref{model}) and deduced its $T=0$
phase-diagram in the model parameter space. 
\enlargethispage*{100cm}
\pagebreak
The Hamiltonian is given by 
\begin{equation}
 \label{model_ham}
{H}=J_1\underbrace{\sum_{<i,j>} \vec S_i\cdot \vec S_j}_{H_0}
+J_2\underbrace{\sum_{<i,k>} \vec S_i\cdot \vec S_k}_{H_1}
+J_3\underbrace{\sum_{<i,l>} \vec S_i\cdot \vec S_l}_{H_2},
\end{equation}
where the bonds corresponding to
interactions $J_1$, $J_2$ and $J_3$ are shown in Fig.~\ref{model}.
For $J_3=0$ the model reduces to the original Shastry-Sutherland
model.\\
It can be easily verified that the singlet-dimer state
(singlets on all strong bonds $J_1$, henceforth called dimers) is 
an exact eigen state of our model: a single spin interacts
via $J_2$ and $J_3$ with $S=0$ objects only. Thus the corresponding
terms in Hamiltonian~(\ref{model_ham}) do not contribute. The remaining
expression simply gives $-(3/8) J_1$ per spin. This article focuses on
the region, where the singlet-dimer state is the ground state, called
the dimer phase. In this phase a single excitation
(magnon) is introduced by breaking up one singlet and substituting one
triplet instead (the triplet's z-component is irrelevant). By hopping
from dimer to dimer this triplet acquires a
dispersion, which we intend to calculate. It suffices to do the
calculations on an effective square lattice $\Gamma_{\rm eff}$, where one
site represents one dimer. Note that $\Gamma_{\rm eff}$ has a
sublattice structure A/B, where A (B) corresponds to vertical
(horizontal) dimers.\\
We derive a power series 
expansion about the limit of strong dimerization ($J_2, J_3
\rightarrow 0$) for the one-magnon
dispersion. By fitting the expansion to
INS-Data obtained by Kageyama {\it et al.} \cite{kagey00} we deduce
model parameters which show good agreement with parameters previously
determined. \cite{knett00b}\\
The next section gives a short introduction to the method
used. Before we calculate and discuss the one-magnon dispersion
(Sec.~\ref{sec_disp}) we use some of the model's symmetries 
to derive interesting and useful relations between various
hopping amplitudes (Sec.~\ref{Symm}).

\section{Method and qualitative pictures}
\label{method}
The one-magnon dispersion is calculated perturbatively about the
limit of isolated dimers using the flow equation
method introduced previously \cite{knett99a}. Given a perturbation
problem that can be formulated in the standard way 
\begin{equation}
  \label{h_pert}
  {\cal H}={\cal H}_0 + x{\cal H}_{\rm S}
\end{equation}
this method in its present formulation relies only on two further prerequisites:
\begin{enumerate}[(A)]
\item The unperturbed Hamiltonian ${\cal H}_0$ must have an equidistant
  spectrum bounded from below. The difference between two successive
  levels is called an energy quantum.
\item There is a number ${\mathbb N} \ni N>0$ such that the perturbing Hamiltonian
  $H_{\rm S}$ can be written as $H_{\rm S}=\sum_{n=-N}^{N}T_n$ where
  $T_n$ increments (or decrements, if $n<0$) the number of energy
  quanta by $n$.
\end{enumerate}

The flow equation method maps the perturbed
Hamiltonian ${\cal H}$ by a continuous unitary transformation to an effective
Hamiltonian ${\cal H}_{\rm eff}$, which {\it conserves} the number of
energy quanta, i.e. $[{\cal H}_{\rm eff},{\cal H}_0]=0$. Thus the effective
Hamiltonian is block diagonal and has the form 
\begin{equation}
\label{H_eff}
{\cal H}_{\rm eff} = {\cal H}_0 +\sum_{k=1}^{\infty}x^{k} 
\sum_{|\underline{m}|=k, M(\underline{m})=0} C(\underline{m}) 
T(\underline{m})\ ,
\end{equation}
where $\underline{m}$ is a vector of dimension $k$ of which the
components are in $\{\pm N, \pm (N-1),\ldots \pm 1, 0\}$; $M(\underline{m})=0$
signifies 
that the sum of the components vanishes which reflects the conservation of the
number of energy quanta. The operator product
$T(\underline{m})$ is defined by
$T_{\underline{m}}=T_{m_1}T_{m_2}\cdots T_{m_k}$, where $k$ is the
order of the process. The coefficients 
$C(\underline{m})$ are generally valid fractions, which we computed up
to order $k=15$ for $N=1$ and up to order $k=10$ for $N=2$ (cf. Ref. \cite{knett99a}).\\
We want to emphasize that the effective Hamiltonian ${\cal H}_{\rm
  eff}$ with known coefficients $C(\underline{m})$ 
can be used straightforwardly in all perturbative problems that meet
conditions (A) and (B). The interested reader can find  the
$C(\underline{m})$ and additional information on
our home-pages. For checking purposes we tested the $C(\underline{m})$ by
applying the method to toy models. In the $N=1$ case, for instance, we
considered the one dimensional harmonic oscillator perturbed by itself
and an additional linear potential. Thus, with $\hbar\omega=1$
\begin{eqnarray}
  \label{toy_1}
  {\cal
    H}_0&=&\frac{1}{2}(P^2+X^2)=a^{\dagger}a+\frac{1}{2}\\\label{toy_2} 
  {\cal H}_{\rm
    S}&=&\frac{1}{2}(P^2+X^2)+X=\underbrace{a^{\dagger}a+\frac{1}{2}}_{T_0}+ 
  \underbrace{\frac{1}{\sqrt 2}a^{\dagger}}_{T_1}+ 
  \underbrace{\frac{1}{\sqrt 2}a}_{T_{-1}}\ .
\end{eqnarray}
In this case Eq.~(\ref{h_pert}) can be solved exactly to give
\begin{equation}
  \label{E_n}
  E_n=(1+x)\left(n+\frac{1}{2}\right)-\frac{1}{2}\frac{x^2}{1+x}\ .
\end{equation}
We expand this last equation about $x=0$ and obtain
\begin{equation}
  \label{E_n_taylor}
  E_n=n+\frac{1}{2}+\frac{1}{2}(2n+1)x-\frac{1}{2}\sum_{i=2}^{\infty}(-1)^{i}x^i\ . 
\end{equation}
By inserting the $T_i$ defined in Eq.~(\ref{toy_2}) in the effective
Hamiltonian~(\ref{H_eff}) and calculating $\langle n|{\cal H}_{\rm
  eff}|n\rangle$ with $n\in \mathbb{N}$ we retain Eq.~(\ref{E_n_taylor})
exactly up to 15$^{\rm th}$ order.\\

\begin{figure}[ht]
\begin{center}
{\psfig{figure=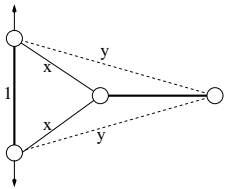 ,height=25mm}}
\end{center}
\vspace*{-1mm}
\caption{\footnotesize{A pair of adjacent dimers (intra dimer coupling constant
    set to unity) connected by the perturbing interactions $x$ and
    $y$. The local operators ${\mathcal{T}}_{i}$ as 
    defined in Tab.~\ref{t_tab} acquire a global minus, if we reflect
    the pair on the indicated axis and keep the notation
    $|x_1,x_2\rangle$ as defined in the text. This is due to the
    singlet antisymmetry under reflection.} \label{el_dim}}
\end{figure}

We now show that Hamiltonian~(\ref{model_ham}) meets conditions (A)
and (B) for $N=1$. To this end  we
rewrite Eq.~(\ref{model_ham}) 
\begin{equation}
  \label{H_pert}
  \frac{H}{J_1} = H_0 + x H_{\rm S}\ ,\ \ \mbox{with}
\end{equation}
\begin{equation}
  \label{H_S}
  H_{\rm S} = H_1 + \frac{y}{x} H_2\ ,\ \  x=\frac{J_2}{J_1}\ \mbox{  and
  }\ y=\frac{J_3}{J_1}\ .
\end{equation}
In the limit of isolated dimers ($x=0$, with $y/x$
finite) $H$ is bounded from below and has an equidistant energy
spectrum since up to a trivial constant $H_0$ simply counts the number of
excited dimers, i.e. energy quanta.\\
To decompose $H_{\rm S}$ we follow the same proceeding as in
Ref.~\cite{knett99a} and state the result
\begin{equation}
  \label{H_S_decomp_1}
  H_{\rm S}=T_{-1}+T_{0}+T_1
\end{equation}
with
\begin{eqnarray}
  \label{H_S_decomp_2}
  T_{\pm 1}&=&\frac{1}{2}\left( 1-\frac{y}{x} \right)\sum_{\nu} {\mathcal
    T}_{\pm 1}(\nu)\\\label{H_S_decomp_3}
  T_{0}&=&\frac{1}{2}\left( 1+\frac{y}{x} \right)\sum_{\nu} {\mathcal
    T}_{0}(\nu)\ ,
\end{eqnarray}
where $\nu$ denotes the pairs of adjacent dimers. For some fixed $\nu$ we
encounter the pair depicted in Fig.~\ref{el_dim}, the state of which 
is determined by $|x_1,x_2\rangle$, where $x_1, x_2 \in
\{s,t^1,t^0,t^{-1}\}$ are singlets or one of the
triplets occupying the vertical (horizontal) dimer,
respectively. The superscript $n\in \{0,\pm1\}$ in $t^n$ stands for the
$S^z$ component. For the pair $|x_1,x_2\rangle$ we give the action of the local
operators ${\mathcal{T}}_{i}$ in Tab.~\ref{t_tab}.
\begin{table}[ht]
\begin{center}
\begin{tabular}{|ccc|}
\hline
&${\mathcal{T}}_{0}$&\\
\hline\hline
$|t^{\pm 1},t^{\pm 1}\rangle$ & $\longrightarrow$ & 
$|t^{\pm 1},t^{\pm 1}\rangle$\\
$|t^{\pm 1},t^{0}\rangle$ & $\longrightarrow$ & 
$|t^{0},t^{\pm 1}\rangle$\\
$|t^{\pm 1},t^{\mp 1}\rangle$ & $\longrightarrow$ & 
$|t^{0},t^{0}\rangle-|t^{\pm 1},t^{\mp 1}\rangle$\\
$|t^{0},t^{0}\rangle$ & $\longrightarrow$ & 
$|t^{1},t^{-1}\rangle+|t^{-1},t^{1}\rangle$\\
$|t^{0},t^{\pm 1}\rangle$ & $\longrightarrow$ & 
$|t^{\pm 1},t^{0}\rangle$\\
\hline
\hline
&${\mathcal{T}}_{ 1}$&\\
\hline\hline
$|t^{\pm 1},s\rangle$ & $\longrightarrow$ & 
$\mp |t^{0},t^{\pm 1}\rangle \pm |t^{\pm 1},t^0\rangle$\\
$|t^{0},s\rangle$ & $\longrightarrow$ & 
$|t^{1},t^{-1}\rangle-|t^{-1},t^1\rangle$\\
\hline
\end{tabular}
\end{center}
{\caption {\footnotesize The action of the local operators
    ${\mathcal{T}}_{0}$ and ${\mathcal{T}}_{1}$ as they appear in
    Eqs.~(\ref{H_S_decomp_2},\ref{H_S_decomp_3}) on all
    relevant states of the dimer pair depicted  in
    Fig.~\ref{el_dim}. These operators conserve the total $S^z$
    component. Note that ${\mathcal{T}}_{1}$ can only create 
    another triplet on the horizontal dimer if there exists already 
    one on the vertical dimer. This has also been noticed in
    Ref.~[10]. Possible 
    ${\mathcal{T}}_{\pm 2}$ operators cancel out due to the lattice inherent
    frustration. Matrixelements not listed are zero.} \label{t_tab}}
\end{table}
The remaining matrix elements can be constructed by using ${\mathcal
  T}_{n}^{\dagger} = {\mathcal T}_{-n}$. Note that we need to fix the
orientation for singlets, say spin up with positive sign always at
the right (upper) site of the dimers. Hence ${\mathcal
  T}_{1}$ and ${\mathcal T}_{-1}$ acquire a global minus for oppositely
oriented dimer pairs (reflection of the dimer pair in
Fig.~\ref{el_dim} about the vertical dimer).

Instead of
Hamiltonian~(\ref{model_ham}) we can now  
use the effective Hamiltonian ${\cal H}_{\rm eff}$
(Eq.~(\ref{H_eff})) with the $T_i$ defined in
Eqs.~(\ref{H_S_decomp_2},\ref{H_S_decomp_3}).
The effective Hamiltonian~(\ref{H_eff}) simplifies the
computations considerably. Since $[{\cal H}_{\rm eff},{\cal H}_0]=0$, 
${\cal H}_{\rm eff}$ is block diagonal
allowing triplet conserving processes only. Thus the effective
Hamiltonian acts in a much smaller Hilbert space than the original
problem, which is a great advantage in the numerical implementation.\\
In addition to this simplification the explicit form of ${\cal
  H}_{\rm eff}$ provides a simple and comprehensive picture of the 
physics involved. Imagine we were to put a triplet on one of the
dimers in the lattice. In the real substance this local excitation
would polarize its environment due to the exchange couplings and
must be viewed as a dressed quasi particle surrounded by a cloud of
virtual excitations fluctuating in space and
time. The effective Hamiltonian includes these fluctuations as virtual
processes 
$T(\underline{m})$, each weighted by the factor
$C(\underline{m})$. Each process ends with a state having the same
number of triplets as the initial state. Other quantum numbers
such as the total spin are also conserved. As the
order $k$ is increased more 
and longer processes are allowed for and the accuracy of the
results will be enhanced. Inspecting the weight factors
$C(\underline{m})$ shows that longer processes have less influence. 

\begin{figure}[ht]
\begin{center}
{\psfig{figure=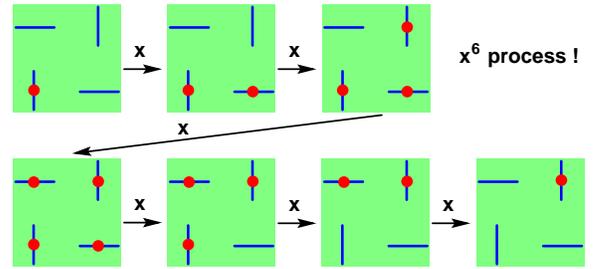 ,height=35mm}}
\end{center}
\vspace*{-1mm}
\caption{\footnotesize{Leading (virtual) process for one-triplet
    hopping corresponding to
    $x^6T(\underline{m})=x^6T_{-1}T_{-1}T_{-1}T_1T_1T_1$. Dark dots are
    triplets, bars are dimers. There is no lower order process leading
    to one-triplet motion.} \label{hopp}}
\end{figure}

Let us follow one of the possible virtual processes and understand why
the observed triplet dispersion of 
SrCu$_2$(BO$_3$)$_2$ is rather flat (cf. Fig.~\ref{ins}). Suppose
we begin with one triplet in the lattice as depicted in the upper left
corner of Fig.~\ref{hopp}.
By applying $T_1$ once we can create another triplet only on one of
the two $horizontally$ adjacent dimers as is clear from Tab.~\ref{t_tab}
and Fig.~\ref{el_dim}. From there we 
might create another one and so on till we can close a circle
(bottom left state in Fig.~\ref{hopp}). We now start destructing
triplets by $T_{-1}$ processes and end up with the shifted triplet. The
amplitude for this hopping is $\propto x^6$. It is the largest
amplitude one can find (see also Ref.~\cite{miyah00a}). Therefore the 
triplets are rather localized leading to a flat dispersion.

Note that the $T_1$ ($T_{-1}$) processes are proportional to $(x-y)$
(see Eq.~(\ref{H_S_decomp_2})). So, the leading triplet motion is
${\cal O}((x-y)^6)$ in leading order. Even local processes (without
hopping) include at least two $T_1$ ($T_{-1}$) processes. Hence all
perturbative amplitudes are at least of order $(x-y)^2$.
We will make use of this fact later on.

Examining the two-triplet sector~\cite{knett00b} we showed that
$correlated$ hopping processes occur in second order already. The
actual dispersion, however, sets in only in third order. This much lower order
($x^3$ instead of $x^6$) explains the
much stronger two-magnon dispersion~\cite{kagey00}.

To quantify the picture constructed
let $|{\mathbf r}\rangle = 
|r_1,r_2\rangle$ denote the state of the 
system with one triplet at ${\mathbf r} \in \Gamma_{\rm eff}$ and
singlets on all other sites. The amplitude $t_{{\mathbf r'}- {\mathbf
    r}}^{o({\mathbf r})}$ for a 
triplet-hopping from site ${\mathbf r}$  to site ${\mathbf r'}$ is 
given by 
\begin{equation}
\label{hopp_def}
t_{{\mathbf r'}-{\mathbf r}}^{o({\mathbf r})}=\langle{\mathbf r'}|{\cal H}_{\rm eff}|{\mathbf r}\rangle\ ,
\end{equation}
where the upper index $o({\mathbf r}) \in \{v,h\}$ allows to
distinguish whether the hopping started on a vertically oriented ($v$)
or a horizontally oriented dimer ($h$). Furthermore we choose to split the
hopping amplitudes into an average part $\bar{t}_{\mathbf s}$ and an
alternating part $dt_{\mathbf s}$ (${\mathbf s}={\mathbf r'}-{\mathbf r}$)
\begin{equation}
  \label{split}
  t^{o({\mathbf r})}_{\mathbf s}=\bar{t}_{\mathbf s}+e^{i{\mathbf Q}{\mathbf
  r}} dt_{\mathbf s}\ ,
\end{equation}
with ${\mathbf Q}=(\pi,\pi)$.\\
The right hand
side of Eq.~(\ref{hopp_def}) can be easily implemented on a
computer. For details see Ref.~\cite{knett99a}. 
We want to point out, however, that all computations are done
symbolically, i.e. we obtain results as functions (polynomials) of all model
parameters.

\section{Symmetries}
\label{Symm}
\begin{figure}[ht]
\begin{center}
{\psfig{figure=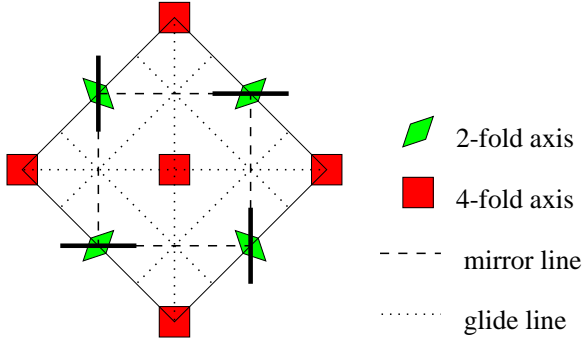 ,height=45mm}}
\end{center}
\vspace*{-1mm}
\caption{\footnotesize{A possible unit cell of the Shastry-Sutherland
    model with the dimers arranged at the sides of the cell. All
    symmetries are depicted. The two-dimensional 
    crystallographic space group is p4mm.} \label{unicell}}
\end{figure}

Before we calculate the one-triplet dispersion quantitatively in the
next section it is worthwhile to look at the symmetries the model in
Fig.~\ref{model} displays. The two-dimensional space group of the model
can be identified to be p4mm with the underlying point group 4mm as
can be verified in Fig.~\ref{unicell}.
We will not treat all symmetry aspects but concentrate on those
that will be of use later on. We choose the coordinate system parallel
to the dimers such that
one dimer (horizontal or vertical) lies in the origin and introduce two
diagonals u and v crossing the origin with slope -1 and 1
, respectively. The distance between the centers of two adjacent dimers
is set to unity. 

Several relations between different hopping amplitudes can be
deduced. To this end we define six symmetry operations which map 
the lattice onto itself. Note that the fixed singlet orientation can
lead to negative phase factors.\vspace*{2mm}\\
$m_{x/y}$: Reflection about x/y-axis\\
$m_{x/y}|r_1,r_2\rangle =
(-1)^{r_1+r_2}|\pm r_1,\mp r_2\rangle$.\vspace*{2mm}\\
$I$: Inversion about the origin\\
$I|r_1,r_2\rangle =
|-r_1,-r_2\rangle$.\vspace*{2mm}\\
$\sigma_{v/u}$: Reflection about the diagonals v/u plus translation by $(0,-1)$\\
$\sigma_{v/u}|r_1, r_2\rangle =
|\pm r_2,\pm r_1-1\rangle$.\vspace*{2mm}\\
$R$: Rotation of $\pi/2$ about the origin plus translation by
$(0,-1)$\\
$R|r_1,r_2\rangle =
(-1)^{r_1+r_2}|-r_2,r_1-1\rangle$.\vspace*{0mm}\\

Since these operations leave ${\cal H}$ unchanged they all commute with ${\cal H}_{\rm eff}$.
Applying $m_x$ to Eq.~(\ref{hopp_def}) we find
\begin{eqnarray}
  \nonumber
  t^{o({\mathbf r})}_{{\mathbf r'}-{\mathbf r}}&=&(-1)^{r_1'+r_2'}\langle
  r_1',-r_2'|m_x{\cal H}_{\rm eff}|r_1,r_2\rangle\\\nonumber
  &=&(-1)^{r_1+r_2+r_1'+r_2'}\langle r_1',-r_2'|{\cal H}_{\rm
  eff}|r_1,-r_2\rangle\\\label{m_x_l}
  &=&(-1)^{r_1+r_2+r_1'+r_2'}t^{o({\mathbf
  r})}_{(r_1',-r_2')-(r_1,-r_2)}\ ,
\end{eqnarray}
or simply
\begin{equation}
  \label{m_x}
  t^{o({\mathbf r})}_{\mathbf s}=(-1)^{s_1+s_2}t^{o({\mathbf
  r})}_{s_1, -s_2}\ .
\end{equation}
Analogously using $m_y$, $I$, $\sigma_u$, $\sigma_v$ and $R$
we find
\begin{eqnarray}
  \label{m_y}
  t^{o({\mathbf r})}_{\mathbf s}&=&(-1)^{s_1+s_2}t^{o({\mathbf
  r})}_{-s_1, s_2}\\
\label{I}
&=&t^{o({\mathbf r})}_{-{\mathbf s}}\\
\label{s_u}
&=&t^{o({\mathbf r}-(0,1))}_{-s_2, -s_1}\\
\label{s_v}
&=&t^{o({\mathbf r}-(0,1))}_{s_2, s_1}\\
\label{R}
&=&(-1)^{s_1+s_2}t^{o({\mathbf r}-(0,1))}_{-s_2, s_1}
\end{eqnarray}
respectively.
In particular Eqs.~(\ref{m_y}) and (\ref{I}) yield together
\begin{eqnarray}
  \label{vanish}
  \nonumber
  t_{(-s_1,0)}^{o({\mathbf r})}&=&(-1)^{s1}t_{(s1,0)}^{o({\mathbf r})}=t_{(s1,0)}^{o({\mathbf r})}\\
  \Rightarrow t_{(s_1,0)}^{o({\mathbf r})}&=&0\mbox{, if } s_1 \mbox{ odd,
  analogously}\\
\Rightarrow t_{(0,s_2)}^{o({\mathbf r})}&=&0\mbox{, if } s_2 \mbox{ odd,}
\end{eqnarray}
describing the interesting fact, that hopping along the axis has non zero
amplitude only if this hopping moves the triplet an even number of
sites ($\Gamma_{\rm eff}$).

\section{Dispersion}
\label{sec_disp}
Since ${\cal H}_{\rm eff}$ conserves the number of triplets the
one-triplet dispersion is particularly easy to calculate. Starting with
one triplet this excitation can only be shifted, i.e. the triplet hops
on the effective lattice $\Gamma_{\rm eff}$. Additional care has to be
taken to account for the sublattice structure of $\Gamma_{\rm
  eff}$. From Eq.~(\ref{hopp_def}) we get
\begin{equation}
  \label{H-eff_k}
  {\cal H}_{\rm eff}|{\mathbf r}\rangle = \sum_{\mathbf r'}t^{o({\mathbf
  r})}_{{\mathbf r'}}|{\mathbf r}+{\mathbf r'}\rangle\ .  
\end{equation}
We introduce Fourier transformed states
\begin{equation}
  \label{k_zust}
  |\sigma,{\mathbf k}\rangle=\frac{1}{{\sqrt L}}\sum_{\mathbf r}|{\mathbf r}\rangle
  e^{i({\mathbf k}+\sigma Q) {\mathbf r}}
\end{equation}
with the number of dimers $L$, the new quantum number $\sigma \in
\{0,1\}$ reflecting the sublattice structure and ${\mathbf k}$ a
vector in the magnetic Brillouin zone (MBZ). Calculating the action
of ${\cal H}_{\rm eff}|\sigma,{\mathbf k}\rangle$ on these states
yields
\begin{eqnarray}
  \nonumber
  {\cal H}_{\rm eff}|\sigma,{\mathbf k}\rangle &=&
  \frac{1}{\sqrt{L}}\sum_{{\mathbf r},{\mathbf r'}}(\bar{t}_{\mathbf
  r'}+e^{iQ{\mathbf r}}dt_{\mathbf r'})|{\mathbf
  r}+{\mathbf r'}\rangle e^{i({\mathbf k}+\sigma Q){\mathbf r}}\\\nonumber
&=& \sum_{\mathbf r'}\bar{t}_{\mathbf
  r'}e^{-i({\mathbf k}+\sigma Q) {\mathbf
  r'}}\underbrace{\frac{1}{\sqrt{L}}\sum_{\mathbf r}|{\mathbf r}\rangle
  e^{i({\mathbf k}+\sigma Q){\mathbf r}}}_{|\sigma,{\mathbf k}\rangle}\\\label{cal_1}
&&\hspace*{-19mm} +\sum_{\mathbf r'}dt_{\mathbf
  r'}e^{-i({\mathbf k}+\bar{\sigma} Q) {\mathbf
  r'}}\underbrace{\frac{1}{\sqrt{L}}\sum_{\mathbf r}|{\mathbf r}+{\mathbf r'}\rangle
  e^{i({\mathbf k}+\bar{\sigma} Q)({\mathbf r}+{\mathbf
  r'})}}_{|\bar{\sigma},{\mathbf k}\rangle}\ ,
\end{eqnarray}
with $\bar{\sigma}=1-\sigma$. We used the definitions (\ref{hopp_def}) and
(\ref{split}). Further, since Eq.~(\ref{I}) holds for
even and odd $r_1+r_2$ one has $\bar{t}_{\mathbf
  s}=\bar{t}_{-{\mathbf s}}\mbox{\ and\ }dt_{{\mathbf s}}=dt_{-{\mathbf s}}$.
Thus we can simplify the sums over ${\mathbf r'}$ in
Eq.~(\ref{cal_1}) to give
\begin{eqnarray}
  \nonumber
  {\cal H}_{\rm eff}|\sigma,{\mathbf k}\rangle &=&
  \underbrace{\left[\bar{t}_{\mathbf 0}+2\sum_{{\mathbf r} > 0}\bar{t}_{\mathbf
  r}\cos(({\mathbf k}+\sigma Q){\mathbf r}) \right]}_{a_{\sigma}}|\sigma,{\mathbf
  k}\rangle \\\label{2x2}
&+&
  \underbrace{\left[dt_{\mathbf 0}+2\sum_{{\mathbf r} > 0}dt_{\mathbf
  r}\cos(({\mathbf k}+\bar{\sigma}Q){\mathbf r}) \right]}_{b}|\bar{\sigma},{\mathbf
  k}\rangle\ ,
\end{eqnarray}
with ${\mathbf r}>0$ if and only if ($r_1>0$ or $r_1=0$ but $r_2>0$). In Appendix~A
we show that $dt_{\mathbf 0}=0$ and $dt_{\mathbf r}=0$ for $r_1+r_2$
odd. Hence $b$ does not
depend on $\bar{\sigma}$
\begin{equation}
  \label{b-b}
  b=2\!\!\!\!\!\!\sum_{{\mathbf r} > 0 \atop r_1+r_2\mbox{ \scriptsize even}}\!\!\!\!\!\!dt_{\mathbf r}\cos({\mathbf k}{\mathbf r})\ ,
\end{equation}
and ${\cal H}_{\rm eff}$ is symmetric in the new states.
The remaining $2\times2$ problem can be solved easily to give the dispersion
\begin{equation}
  \label{disp}
  \omega_{1/2}({\mathbf
  k})=\underbrace{\frac{a_0+a_1}{2}}_{\omega_0({\mathbf k})}\pm
  \frac{1}{2}\sqrt{(a_0-a_1)^2+4b^2}\ .
\end{equation}
Thus the one-triplet dispersion splits into two branches. We want to
point out, however, that at ${\mathbf k}={\mathbf 0}$ and at the
borders of the MBZ (i.e. $|k_x+k_y|=\pi$ or $|k_y-k_x|=\pi$) the two
branches fall onto each other leading to a 
two-fold degenerate dispersion. An analogous degeneracy is noticed 
in the two-triplet sector \cite{knett00b}. In
Appendix~B we demonstrate that the degeneracy is due to the glide line symmetries $R$ and
$\sigma_{u/v}$ and show that $(a_0-a_1)$ and $b$ both
vanish. Moreover, $(a_0+a_1)$ is a sum over ${\mathbf r}$ with
$r_1+r_2$ even and thus contains ``even" hopping amplitudes only. At
${\mathbf k}={\mathbf 0}$ and at the  border 
of the MBZ we therefore find that one-triplet hopping takes place 
on one species of the two sublattices A/B in $\Gamma_{\rm eff}$
only. In other words, the triplets live either on the horizontal or on
the vertical dimers.

We calculated the amplitudes $\bar{t}_{\mathbf r}$ and $dt_{\mathbf
  r}$ (and therefore the dispersion) as exact polynomials in $x$ and
$y$ up to and including 15$^{\rm th}$ order. On appearance of this article
  these polynomials will be published in 
electronic form on our home pages.\\
Expanding the square root in Eq.~(\ref{disp}) about the limit of
vanishing $x$ and $y$ produces terms $\propto x^{\alpha}y^{\beta}$ with
$\alpha+\beta \ge 10$. Hence the energy splitting starts in
10$^{\rm th}$ order and is negligible for all reasonable values of $x$
and $y$. The fact that the splitting starts four orders later
than the dispersion may be understood by observing that ($a_0-a_1$)
is a sum over ${\mathbf r}$ with $r_1+r_2$ odd. From the discussion
at the end of section \ref{Symm} it is clear that $t_{\pm
  1,0}^{o({\mathbf r})}$ and $t_{0, \pm 1}^{o({\mathbf r})}$ vanish so
  that the leading ``odd"
process is $t_{\pm 2,\pm 1}^{o(r)}$ or $t_{\pm 1,\pm 2}^{o(r)}$, which
  start in 10$^{\rm th}$ order only. 
One may object that $b$ contains $dt_{1,1}$, which could be larger,
but from relation (\ref{s_s_1_s_2}) in Appendix A follows that
  $dt_{1,1}$=0. The amplitudes $dt_{\pm 2,0}$ and $dt_{0,\pm 2}$ in
  $b$ also start in 10$^{\rm th}$ order only. Hence the almost
  degeneracy in the one triplet sector can be understood on the basis
  of the symmetries of the lattice.

By substituting $y=0$ in $\omega_0({\mathbf k})$ we obtain the
decimal numbers computed by Zheng et al. \cite{weiho98b}. Our series
expansion for the dispersion (\ref{disp}) converges nicely 
but it can be improved further by the use of D-log Pad\'e
approximants~\cite{domb83_pade}. 

\begin{figure}[ht]
\begin{center}
{\psfig{figure=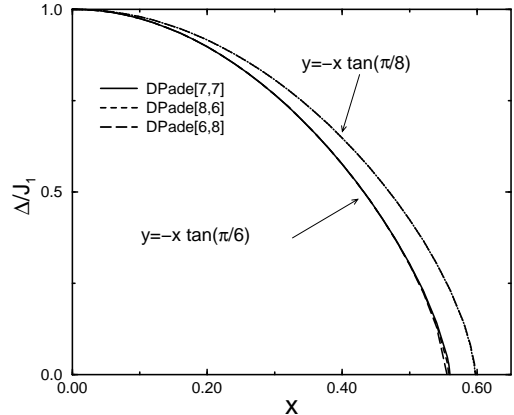 ,height=65mm}}
\end{center}
\vspace*{-1mm}
\caption
{
  \footnotesize
  {
    D-log Pad\'e extrapolations of the gap along two lines
    in the ($x,y$)-plane. Different approximants for a fixed ratio
    coincide very well. The ($x$,$y$) values where the gap vanishes
    constitute a line in the ($x,y$)-plane indicating the definitive
    collapse of the dimer phase (cf. Fig.~2 in Ref.~[6]).
    }
  \label{dpade_bild}
  }
\end{figure}

\begin{figure}[ht]
\begin{center}
{\psfig{figure=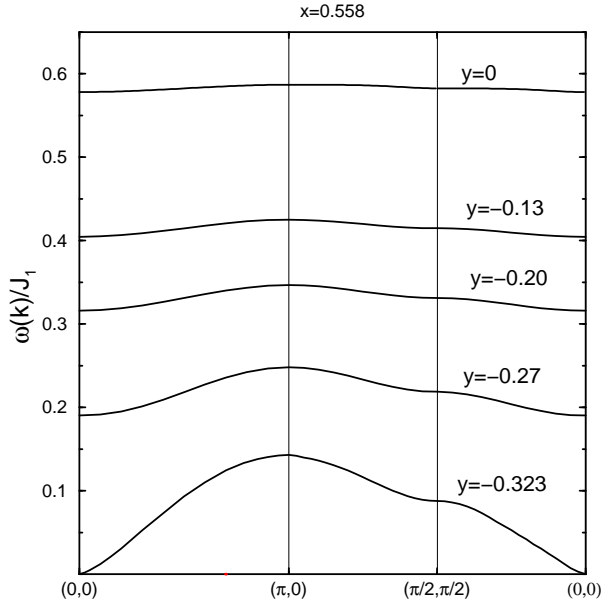 ,height=75mm}}
\end{center}
\vspace*{-1mm}
\caption{\footnotesize{One-triplet dispersion
    in the MBZ at various points of
    the ($x$,$y$)-plane. The lowest curve shows $\omega({\mathbf k})$ at
    the point where the gap vanishes. With increasing $y$ the
    dispersion decreases considerably as is clear from $(\omega(\pi
    ,0)-\omega(0,0))\propto (x-y)^6$. } \label{disp1}} 
\end{figure}

 Let us first consider the energy gap
$\Delta:=\omega(0,0)$ as function of $x$ and $y$. We fix the ratio of
$x$ and $y$ and use D-log Pad\'e approximants to extrapolate the
remaining expression. For all $x/y$ ratios we tested, we
find the approximants to be very stable, i.e. most of the possible
approximants at a fixed ratio coincide very well. (Some are defective, i.e they
show spurious singularities.) Fig.~\ref{dpade_bild} shows D-log Pad\'e
extrapolations for $y=-x\tan(\pi/6)$ and 
$y=-x\tan(\pi/8)$. In Ref.~\cite{mulle00} we used this technique
to determine the line in the ($x,y$)-plane where the gap $\Delta$
vanishes. The vanishing of $\Delta$ indicates definitively the break
down of the dimer phase. But it may happen that another excitation becomes
soft before the elementary triplet vanishes (cf. Ref.~\cite{knett00b})
or that a first order transition takes place
\cite{knett00b,albre96}. In the present work we choose to examine the
elementary triplet only.
For $y=-x\tan(\pi/6)$ the gap vanishes
at $x=0.558(1)$ ($y=-0.323(1)$). At this point we get the lowest curve
in Fig.~\ref{disp1} where we choose to show the dispersion along a
triangle in the MBZ. 
In Sec.~\ref{method} we saw that all hopping amplitudes are
proportional to $(x-y)^6$. Since the dispersion is a sum over these
amplitudes it is clear that $(\omega(\pi, 0)-\omega(0,0))\propto
(x-y)^6$. Indeed, for increasing $y$ we see that the dispersion
decreases 
(Fig.~\ref{disp1}). At $x=y$ we find the dispersion to be
absolutely flat. This is a signiture of the fact that at
$x=y$ the total spin on each $J_1$ bond is a conserved quantity
\cite{mulle00}. Thus there will be no triplet motion.

\begin{figure}[ht]
\begin{center}
{\psfig{figure=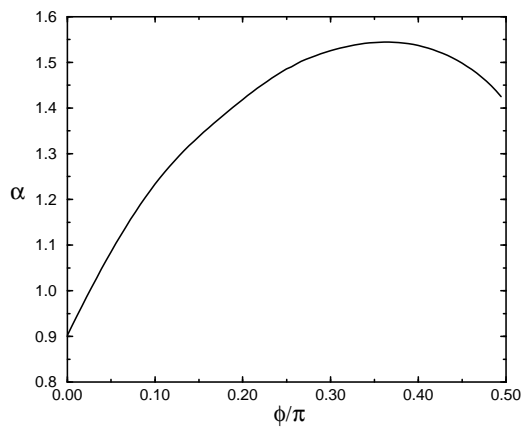,height=65mm}}
\end{center}
\vspace*{-1mm}
\caption{\footnotesize{The exponent $\alpha$ in $\omega(k_x)\propto
    k_x^{\alpha}$ as function of the angle $\phi$ between the $x$ and
    ($-y$) axis.} \label{a_f}}
\end{figure}

We turn to the behaviour of the dispersion
$\omega({\mathbf k})$ for small $|{\mathbf k}|$ on the line in the
($x,y$)-plane where the gap vanishes. After fixing the ratio $x/y$ in
Eq.~(\ref{disp}) and applying the D-log Pad\'e technique we end up with
an expression
\begin{equation}
  \label{dpade}
  \omega(x;{\mathbf k})=\exp\Bigg[\int_0^{x}
  \underbrace{A({\mathbf k})\frac{P(x;{\mathbf
  k})}{Q(x;{\mathbf k})}}_{f(x;{\mathbf k})}\Bigg] dx\ ,
\end{equation}
where $A$ is a function of ${\mathbf k}$ and $P$ and $Q$ are
polynomials in $x$ (where the leading coefficient is unity) of order $M$
and $N$, respectively (shorthand: [M,N]-approximant). At ${\mathbf
  k}=0$ the smallest
positive zero of $Q$, say $x_0$, is the point where the exponent
diverges logarithmically to $-\infty$, hence $\omega(x_0;0)=0$. We set
$k_y=0$ and will show that for small $k_x$
\begin{equation}
  \label{w_ans}
\omega(x_0;k_x) \propto k_x^{\alpha}\mbox{ + higher orders}\ .
\end{equation}
Differentiating $\ln \omega$ yields for $k_x \rightarrow 0$
\begin{equation}
  \label{expon}
  \alpha = \lim_{k_x \rightarrow 0} k_x \int_0^{x_0} \partial_{k_x}
  f(x;k_x) dx\ .
\end{equation}
We decompose $P$ and $Q$ in linear factors
\begin{eqnarray}
  \label{lin_fac}
  P(x;k_x)&=&(x-p_1(k_x))(x-p_2(k_x))\cdots(x-p_M(k_x))\\
  Q(x;k_x)&=&(x-q_1(k_x))(x-q_2(k_x))\cdots(x-q_N(k_x))\ ,
\end{eqnarray}
such that $q_1(k_x)|_{k_x=0}=x_0$. In this way the factor $(x-q_1(k_x))$
dominates the behaviour of $\alpha$ for small $k_x$. Further, since
the dispersion is invariant under the substitution $k_x \rightarrow
-k_x$ we have $A,p_i,q_i \propto k_x^2$ for small $k_x$. Thus we are
led to write $q_1(k_x)=x_0+\beta k_x^2$, with positive $\beta$. With
these preparations we rewrite Eq.~(\ref{expon})
\begin{eqnarray}
  \label{exp_2}
  \alpha&=&\lim_{k_x \rightarrow 0} k_x \int_0^{x_0} \frac{2\beta
  k_x}{(x\underbrace{-x_0-\beta k_x^2}_{-q_1})^2}
  g(x;k_x) dx\\
  &=&\lim_{k_x \rightarrow 0} 2 \int_0^{\frac{x_0}{\beta k_x^2}}
  \frac{g(x_0-\beta k_x^2;k_x)}{(1+y)^2} dy\ ,
\end{eqnarray}
where we substituted $x=x_0-\beta k_x^2 y$ and $g(x;{\mathbf k})$ for
$A(k_x)\cdot P(x;k_x)\cdot(x-q_1)/Q(x;k_x)$. Taking the limit
yields
\begin{equation}
  \label{exp_3}
  \alpha = 2 g(x_0;0)\int_0^{\infty}\frac{1}{(1+y)^2}dy=2
  g(x_0;0)\ .
\end{equation}
A straightforward calculation shows that
\begin{equation}
  \label{exp_f}
  g(x_0;0)=A(0)\frac{P(x_0;0)}{\partial_x
  Q(x_0,0)}\ ,
\end{equation}
which we can easily calculate. Fig.~\ref{a_f} shows the exponent
$\alpha$ as function of the angle $\phi$ measured from the positive
$x$-axis to the negative $y$-axis. 
There is only one ratio $x/y$ for which $\alpha=1$. At this point the
dispersion vanishes with finite spin wave velocity ($\phi \approx
0.081(1)$ or $x\approx 0.679(1)$, $y\approx -0.055(1)$).

\begin{figure}[ht]
\begin{center}
{\psfig{figure=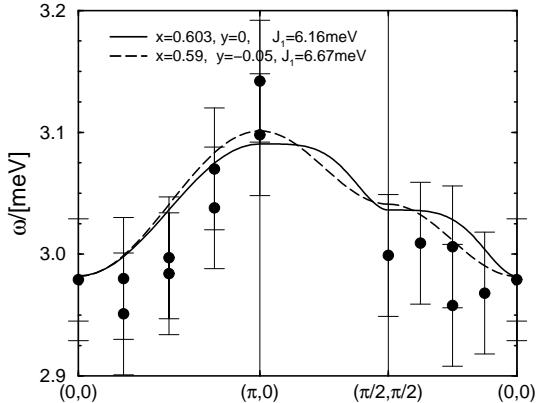 ,height=65mm}}
\end{center}
\vspace*{-1mm}
\caption{\footnotesize{One-triplet dispersion. Our theoretical results
    fitted to INS data (bullets, errors at least as large as error
    bars).  Due to the large errors it is not possible to
    fit the model parameters unambiguously.} \label{ins}}
\end{figure}

Let us turn to a comparison of the theoretical dispersion to
experimental data for SrCu$2$(BO$_3$)$_2$. To fit the dispersion to
experimental data we make use of the parameter 
dependance of our results by requiring the curves to go through
certain points and solve the resulting set of equations. At
${\mathbf k}=(0,0)$ ESR~\cite{nojir99}, FIR~\cite{room99} and INS~\cite{kagey00}
data suggest a value of $\omega(0,0)=2.98$meV. At finite ${\mathbf k}$
we have to rely on the INS measurement, which contain rather large
errors. In Fig.~\ref{ins} we show the INS data (bullets and error bars)
and two of our fitted curves. For the parameter values given we get an
excellent agreement.
Because of the flatness of the dispersion and the comparably large
error bars it is not possible to fix the model
parameters unambiguously . As sketched in Fig.~\ref{ins} one can lower
$x$ and enlarge $J_1$ and $|y|$ without loosing reasonable
agreement. On the basis of the one-triplet dispersion it is not
possible to fix the model parameters quantitatively. Our investigation in the
two-triplet sector~\cite{knett00b} (where $y$ is disregarded, since
it is probably not very important) enables us to fix the
parameters to the intervals ($y=0$,) $x=0.603(3)$ and
$J_1=6.16(10)$meV. The corresponding one-triplet dispersion is the
solid curve in Fig.~\ref{ins}.

\section{Conclusion}
In summary, we have demonstrated the utility of our perturbation
method introduced earlier \cite{knett99a}. By the means of a
continuous unitary transformation the original Hamiltonian is mapped 
onto a block diagonal, energy quanta conserving, effective Hamiltonian
${\cal H}_{\rm eff}$. This 
allows to do the calculations in different energy sectors
separately.\\
Here we used this method to develop a picture of excitation
processes in a model for SrCu$_2$(BO$_3$)$_2$ and calculated the series
expansion of the one-magnon
dispersion $\omega$ about the limit of isolated dimers as a 15$^{\rm th}$ order
polynomial in the model parameters. 

We showed that the dispersion
decomposes into two nearly degenerate branches with a splitting 
proportional to $x^{10}$. At ${\mathbf
  k}={\mathbf 0}$ and at the border of the magnetic Brillouin zone the
two branches fall onto each other. By making use of our detailed
analysis of the models symmetries we showed that 
point group operations alone cannot explain these degenaracies. In
fact we showed that the model's space group symmetries have to be taken
into account.\\
Moreover we analysed the critical behaviour of $\omega$ for small
$|{\mathbf k}|$ at various ratios of $x$ and $y$. It is found that the
dispersion vanishes only for $x\approx 0.679(1)$ and $y\approx
-0.055(1)$ with a finite spin velocity.\\
Finally, we fitted our dispersion to INS data obtained by Kageyama {\it et
  al.} \cite{kagey00} and found that it is not possible to fix the
model parameters unambiguously from the one-triplet data alone due
to very large experimental error 
bars. The parameter values determined by our investigations in the
two-triplet sector (\cite{knett00b}: $y=0$, $x=0.603$ , $J=6.67$meV),
however, lead to a one-magnon
dispersion agreeing nicely with the experimental findings.\\
The authors would like to thank Alexander B\"uhler for helpful
discussions and the Regional Computing Centre of the University of Cologne (RRZK)
and Dr. Kim Yong Taik in particular for generous assistance in our large
scale computations. This work was supported by the DFG-Schwerpunkt
1073 and by the SFB 341.

\appendix
\section{}
\label{dt}
First we show that $dt_{\mathbf r}$=0 if $r_1=r_2$.\\
With Eq.~(\ref{s_v}) (the $\sigma_v$ symmetry) we have
\begin{equation}
  \label{t0}
  t_{\mathbf s}^{o(\mathbf r)}=t_{s_2,s_1}^{o({\mathbf r}-(0,1))}\ .
\end{equation}
Splitting both sides according to Eq.~(\ref{split}) we get
\begin{equation}
  \label{t0_split}
  \bar{t_{\mathbf s}}+e^{iQ{\mathbf r}}dt_{\mathbf s}=
  \bar{t}_{s_2,s_1}+e^{iQ({\mathbf r}-(0,1))}dt_{s_2,s_1}\ ,  
\end{equation}
leading to 
\begin{eqnarray}
  \bar{t_{\mathbf s}}+ dt_{\mathbf s}&=& \bar{t}_{s_2,s_1}-
  dt_{s_2,s_1}\ ,\mbox{\ \ for $r$ even and}\\
\bar{t_{\mathbf s}}- dt_{\mathbf s}&=& \bar{t}_{s_2,s_1}+
  dt_{s_2,s_1}\ ,\mbox{\ \ for $r$ odd.}
\end{eqnarray}
Taking the difference of both equations yields
\begin{equation}
  \label{s_s_1_s_2}
  dt_{s_1,s_2}=-dt_{s_2,s_1}\ ,
\end{equation}
which proves the assertion. In particular $dt_{\mathbf 0}=0$.\\
We now show that $dt_{\mathbf r}=0$ if $r_1+r_2$ is an odd number.\\
According to Eq.~(\ref{hopp_def}) we have
\begin{eqnarray}
\nonumber
t_{{\mathbf r'}}^{o({\mathbf r})}&=&\langle{\mathbf r}+{\mathbf
  r'}|{\cal H}_{\rm eff}|{\mathbf r}\rangle\\\nonumber
&=&\langle{\mathbf r}|{\cal H}_{\rm eff}|{\mathbf r}+{\mathbf
  r'}\rangle=t_{-{\mathbf r'}}^{o({\mathbf r}+{\mathbf r'})}\\
&=&t_{{\mathbf r'}}^{o({\mathbf r}+{\mathbf r'})}
\end{eqnarray}
since ${\cal H}_{\rm eff}$ has only real matrix elements in this basis
(cf. Eq.~(\ref{H_eff})). The last equality follows from
Eq.~(\ref{I}). Splitting both sides according to Eq.~(\ref{split})
yields
\begin{eqnarray}
  \nonumber
\bar{t_{\mathbf r'}}+e^{iQ{\mathbf r}}dt_{\mathbf r'}&=& \bar{t_{\mathbf
  r'}}+e^{iQ{\mathbf r}}e^{iQ{\mathbf r'}}dt_{\mathbf r'}\\\nonumber
\Rightarrow dt_{\mathbf r'}&=&e^{iQ{\mathbf r'}}dt_{\mathbf r'}\\
\Rightarrow dt_{\mathbf r'}&=& 0 \mbox{\ \ \ \ for\ }r_1'+r_2'\mbox{\ odd.}
\end{eqnarray}

\section{Degeneracy of $\omega$}
\label{deg}
The dispersion (\ref{disp}) will be two-fold degenerate if the square
root vanishes. We will use the shorthand ${\mathbf r}$ even/odd for
$r_1+r_2$ even/odd.\vspace*{2mm}\\
I) ${\mathbf k}=0$

A) $b\stackrel{!}{=}0$\\
For ${\mathbf k}=0$ Eq.~(\ref{b-b}) gives
\begin{eqnarray}
  \nonumber
  \!\frac{b}{2}&=&\!\!\!\!\sum_{{\mathbf r} > 0 \atop {\mathbf r}\mbox{ \scriptsize
  even}}\!\!\!\!dt_{\mathbf r}\!=
  \!\!\!\!\sum_{r_1=0 \atop {r_2>0\atop {\mathbf r}\mbox{
  \scriptsize even}}}\!\!\!\!dt_{\mathbf r}
        +\!\!\!\!\sum_{r_1 > 0\atop{ r_2=0\atop {\mathbf r}\mbox{
  \scriptsize even}}}\!\!\!\!dt_{\mathbf r}+
  \!\!\!\!\sum_{r_1 > 0\atop{ r_2 >0\atop {\mathbf r}\mbox{ \scriptsize
  even}}}\!\!\!\!dt_{\mathbf r} 
       + \!\!\!\!\sum_{r_1 > 0\atop{ r_2<0 \atop {\mathbf r}\mbox{
  \scriptsize even}}}\!\!\!\!dt_{\mathbf r}\\\label{summe}
&=& I_1+I_2+I_3+I_4\ .
\end{eqnarray}
We rewrite the first and the third sum on the right hand side of Eq.~(\ref{summe}) 
\begin{equation}
  \label{rewr}
  I_1=\!\!\!\!\sum_{r_1>0 \atop {r_2=0\atop {\mathbf r}\mbox{
  \scriptsize even}}}\!\!\!\!dt_{-r_2,r_1}\mbox{\ and\ }I_3=\!\!\!\!\sum_{r_1>0 \atop {r_2<0\atop {\mathbf r}\mbox{
  \scriptsize even}}}\!\!\!\!dt_{-r_2,r1}\ .
\end{equation}
From Eq.~(\ref{R}) (the $R$ symmetry) we deduce 
\begin{eqnarray}
  \nonumber
  dt_{r_1,r_2}&=&-dt_{-r_2,r_1}\ ,
\end{eqnarray}
and see that $I_1=-I_2$ and $I_3=-I_4$. Hence we find $b=0$.\vspace*{5mm}

B) $(a_0-a_1)\stackrel{!}{=}0$\\
For ${\mathbf k}=0$ we have (cf. Eq.~(\ref{2x2}))
\begin{equation}
  \label{tj}
  \frac{a_0-a_1}{2}=\sum_{{\mathbf r}>0}\bar{t}_{\mathbf r}\left[1-\cos(\pi
  (r_1+r_2))\right]=2\!\!\sum_{{\mathbf r}>0\atop {\mathbf r}\mbox{
  \scriptsize odd}}\bar{t}_{\mathbf r}\ ,
\end{equation}
and by recalling that $dt_{\mathbf r}=0$ for ${\mathbf r}$ odd
Eq.~(\ref{R}) yields
\begin{equation}
  \bar{t}_{r_1,r_2}=-\bar{t}_{-r_2,r_1}\ ,
\end{equation}
so that we can use the same splitting as in
Eq.~(\ref{summe}). Thus the energy-degeneracy at ${\mathbf k}=0$ is
due to the rotational symmetry $R$ as defined in Sec~\ref{Symm}.\vspace*{8mm}\\
II) $k_x+k_y=\pi$

A) $b\stackrel{!}{=}0$\\
Making use of $dt_{\mathbf r}=dt_{-{\mathbf r}}$, which follows from
Eq.~(\ref{I}), we have for $k_y=\pi-k_x$  
\begin{eqnarray}
  \nonumber
  \frac{b}{4}&=&\sum_{{\mathbf r} \atop {\mathbf r}\mbox{ \scriptsize
  even}} dt_{\mathbf r}\cos(k_x(r_1-r_2)+\pi r_2)\\\label{dfg}
&=&\sum_{r_1,r_2 \atop {\mathbf r}\mbox{ \scriptsize
  even}} dt_{r_2,r_1}\cos(k_x(r_2-r_1))\cos(\pi r_1)\ .
\end{eqnarray}
In the last step we choose to rearrange the sum and observe that if
${\mathbf r}$ is even we have $r_1$ and $r_2$ both odd or both
even. In both cases the identity
\begin{equation}
  \label{ide}
  \cos(\pi r_1)=\cos(\pi r_2)
\end{equation}
holds. 
Inserting relation (\ref{s_s_1_s_2}) in the last row of Eq.~(\ref{dfg}) we end up with
 \begin{equation}
  \label{fin}
 \frac{b}{4}=-\sum_{{\mathbf r} \atop {\mathbf r}\mbox{ \scriptsize
  even}} dt_{\mathbf r}\cos(k_x(r_1-r_2))\cos(\pi r_2)\ ,
\end{equation}
resulting in $b=-b$ and thus $b=0$.\vspace*{5mm}

B) $(a_0-a_1)\stackrel{!}{=}0$\\
Analogously to I) B) we have here
\begin{eqnarray}
  \nonumber
  \frac{a_0-a_1}{4}&=&\sum_{\mathbf r}\bar{t}_{\mathbf r}\left[\cos({\mathbf
  k}{\mathbf r})-\cos({\mathbf k}{\mathbf r}+\pi(r_1+r_2))\right]\\\nonumber
&=&2\!\!\!\sum_{{\mathbf r}\mbox{ \scriptsize
  odd}}\bar{t}_{\mathbf r}\cos(k_x(r_1-r_2)+\pi r_2)\\\nonumber
&=&2\!\!\!\sum_{{\mathbf r}\mbox{ \scriptsize
  odd}}\bar{t}_{r_2,r_1}\cos(k_x(r_2-r_1))\cos(\pi r_1)\\\label{eee}
&=&-2\!\!\!\sum_{{\mathbf r}\mbox{ \scriptsize
  odd}}\bar{t}_{r_1,r_2}\cos(k_x(r_1-r_2))\cos(\pi r_2)\ ,
\end{eqnarray}
where the last but one equality follows from Eq.~(\ref{s_v}) (the $\sigma_v$
symmetry), i.e $\bar{t}_{r_1,r_2}=\bar{t}_{r_2,r_1}$, and from the fact
that if ${\mathbf r}$ is odd we have that $r_1$ odd and $r_2$ even or
$r_1$ even and $r_2$ odd. From that we see $\cos(\pi
r_1)=-\cos(\pi r_2)$.\\
The degeneracy over the remaining three borders of the magnetic
Brillouin zone can be shown analogously. It is interesting to note,
that the calculations necessarily involved glide line operations. The
degeneracies can thus not be explained  by considering point group
symmetries only.


\begin{thebibliography}{10}

\bibitem{tanig95}
S. Taniguchi {\it et~al.}, J. Phys. Soc. Jpn. {\bf 64},  2758  (1995).

\bibitem{garre97a}
A.~W. Garrett {\it et~al.}, Phys. Rev. Lett. {\bf 79},  745  (1997).

\bibitem{kagey99}
H. Kageyama {\it et~al.}, Phys. Rev. Lett. {\bf 82},  3168  (1999).

\bibitem{miyah98}
S. Miyahara and K. Ueda, Phys. Rev. Lett. {\bf 82},  3701  (1999).

\bibitem{shast81b}
B.~S. Shastry and B. Sutherland, Physica {\bf 108B},  1069  (1981).

\bibitem{mulle00}
E. M\"uller-Hartmann {\it et~al.}, Phys. Rev. Lett. {\bf 84},  1808  (2000).

\bibitem{kagey00}
H. Kageyama {\it et~al.}, Phys. Rev. Lett. {\bf 84},  5876  (2000).

\bibitem{knett00b}
C. Knetter, A. B\"uhler, E. M\"uller-Hartmann, and G.~S. Uhrig, 
  cond-mat/0005322.

\bibitem{knett99a}
C. Knetter and G.~S. Uhrig, Eur. Phys. J. B {\bf 13},  209  (2000).

\bibitem{miyah00a}
S. Miyahara and K. Ueda, Phys. Rev. B {\bf 61},  3417  (2000).

\bibitem{weiho98b}
Zheng Weihong, C. Hamer, and J. Oitmaa, Phys. Rev. B {\bf 60},  6608  (1999).

\bibitem{domb83_pade}
{\em Phase Transitions and Critical Phenomena}, edited by C. Domb and J.~L.
  Lebowitz (Academic Press, London, 1983), Vol.~13.

\bibitem{albre96}
M. Albrecht and F. Mila, Europhys. Lett. {\bf 34},  145  (1996).

\bibitem{nojir99}
H. Nojiri {\it et~al.}, J. Phys. Soc. Jpn. {\bf 68},  2906  (1999).

\bibitem{room99}
T. Room {\it et~al.}, Phys. Rev. B {\bf 61},  14342  (2000).

\end{thebibliography}

\end{document}